\pdfoutput=1
\documentclass{article}

\usepackage{arxiv}

\usepackage[utf8]{inputenc} 
\usepackage[T1]{fontenc}    

\usepackage[colorlinks,citecolor=black,urlcolor=black,bookmarks=false,hypertexnames=false]{hyperref}  
\usepackage{url}            
\usepackage{booktabs}       
\usepackage{amsfonts}       
\usepackage{nicefrac}       
\usepackage{microtype}      
\usepackage{lipsum}		
\usepackage{graphicx}
\usepackage{doi}
\usepackage{subfigure}
\usepackage{float}
\usepackage{multirow}
\usepackage{wrapfig}
\usepackage{amsmath}

\usepackage[square,numbers]{natbib}

\title{Exploring Data-Driven Chemical SMILES Tokenization Approaches to Identify Key Protein-Ligand Binding Moieties}

\author{Asu Busra Temizer\thanks{These authors contributed equally to this work.}  \\
	Faculty of Pharmacy \\ 
 Department of Pharmaceutical Chemistry\\
	\.{I}stanbul University\\
        \And
        G\"{o}k\c{c}e Uludo\u{g}an$^{*}$ \\
	Department of Computer Engineering\\
	Bogazici University\\
        \And
        R{\i}za \"{O}z\c{c}elik$^{*}$  \\
	Department of Computer Engineering\\
	Bogazici University\\
        \And
        Taha Koulani \\
		Faculty of Pharmacy \\ 
 Department of Pharmaceutical Chemistry\\
	\.{I}stanbul University\\
	\AND
	Elif Ozkirimli \\
	Data and Analytics Chapter\\
	Pharma International Informatics\\
	F. Hoffmann-La Roche AG \\
	\And
	 Kutlu O. Ulgen \\
	 Department of Chemical Engineering \\
    Bogazici University\\
	 \And
	 Nilgün Karalı \\
     Department of Pharmaceutical Chemistry\\
    Istanbul University \\
	 \texttt{karalin@istanbul.edu.tr} \\
	 \And
     Arzucan Özgür \\
	Department of Computer Engineering\\
	Bogazici University\\
	 \texttt{arzucan.ozgur@boun.edu.tr} \\
}

\date{}

\renewcommand{\shorttitle}{Data-Driven SMILES Tokenization to Identify Binding Moieties}

\hypersetup{
pdftitle={Exploring Data-Driven Chemical SMILES Tokenization Approaches to Identify Key Protein-Ligand Binding Moieties},
pdfsubject={q-bio.BM, q-bio.QM, cs.CL, cs.LG, stat.ML},
pdfauthor={Asu Busra Temizer,Gökçe Uludoğan, Rıza Özçelik, Taha 
 Kouhani, Elif Ozkirimli, Kutlu O. Ulgen, Nilgün Karalı, Arzucan Özgür},
pdfkeywords={chemical words, chemoinformatics, machine learning, medicinal chemistry, subword tokenization},
}

\begin{document}
\maketitle

\begin{abstract}
Machine learning models have found numerous successful applications in computational drug discovery. A large body of these models represents molecules as sequences since molecular sequences are easily available, simple, and informative. The sequence-based models often segment molecular sequences into pieces called chemical words (analogous to the words that make up sentences in human languages)  and then apply advanced natural language processing techniques for tasks such as \textit{de novo} drug design,  property prediction, and binding affinity prediction. 
However, the chemical characteristics and significance of these building blocks, chemical words, remain unexplored. This study aims to investigate the chemical vocabularies generated by popular subword tokenization algorithms, namely Byte Pair Encoding (BPE), WordPiece, and Unigram, and identify key chemical words associated with protein-ligand binding. To this end, we build a language-inspired pipeline that treats high affinity ligands of protein targets as documents and selects key chemical words making up those ligands based on tf-idf weighting. Further, we conduct case studies on a number of protein families to analyze the impact of key chemical words on binding.
Through our analysis, we find that these key chemical words are specific to protein targets and correspond to known pharmacophores and functional groups. 
Our findings will help shed light on the chemistry captured by the chemical words, and by machine learning models for drug discovery at large.
\end{abstract}


\section{Introduction}

The last decade has witnessed the rise of machine learning. The models learned to sing \citep{dhariwal2020jukebox}, write \citep{chowdhery2022palm}, and paint \citep{ramesh2022hierarchical} through large unlabeled datasets, despite the absence of well-defined rules for the tasks. Drug discovery is also an excellent application domain for machine learning models. However, unlike images and audio tracks,  chemicals are non-numeric and need intermediate representations upon which machines can learn. 

Text-based representations of chemicals \citep{weininger1988smiles, krenn2020self, o2018deepsmiles} can be used as intermediate representations: they are easily available, simple, and as powerful as more complex representations such as 2D or 3D representations \citep{flam2022language}. The power of the text-based representations in machine learning is partially due to the \textit{chemical language} perspective. The chemical language perspective views text-based representations of chemicals as documents written in a chemical language and borrows advanced approaches from the natural language processing domain to build computational drug discovery models \citep{ozturk2020exploring}. Successful applications of the chemical language perspective in drug discovery include \textit{de novo} drug design \citep{segler2018generating,grechishnikova2021transformer,moret2022perplexity,uludougan2022exploiting}, binding site detection \citep{tsubaki2019compound,deepAffinity,lee2022sequence,gligorijevic2021structure}, and drug-target interaction prediction \citep{shin2019self,abbasi2020deepcda,ozccelik2021chemboost,zhao2022attentiondta}.

At the core of the chemical language perspective are \textit{chemical words}, which correspond to the smaller building blocks of chemicals, similar to the words in natural languages. On the other hand, defining the chemical words poses another research problem since the chemical language, unlike natural (human) languages, has no pre-defined collection of words, \textit{i.e.}, a vocabulary. Several approaches are available to build a chemical vocabulary for state-of-the-art models for downstream tasks such as drug-target affinity prediction, similar compound identification, and \textit{de novo} drug design \cite{huang2020moltrans,ozturk2018novel,li2021smiles}. 
However, to the best of our knowledge, the chemical vocabularies obtained from text based representations are not studied from a chemical perspective so far, and therefore it is unknown whether these chemical vocabularies capture chemical information or not.
This raises the question: do models utilizing chemical words rely on chemically meaningful building blocks, or arbitrary chemical subsequences? Here we seek an answer to this question in order to define the role of these models in new drug design and development from a medicinal chemist's perspective.

Chemical word is an integral concept for the studies that rely on the chemical language hypothesis \cite{ozturk2020exploring}. The chemical language hypothesis interprets the text-based representation (\textit{i.e.} molecular strings) of chemicals as sentences and borrows language processing methods from natural languages. Chemical words are the building blocks of these ``sentences", and, unlike the words in natural languages, need to be discovered. 

Subword tokenization algorithms allow learning the chemical words from large unlabeled corpora of molecular strings. In addition to discovering chemical words, they can segment the chemical sentences into chemical words in the discovered vocabulary. One commonly used algorithm in this context is Byte Pair Encoding (BPE) \cite{sennrich2015neural}, which is widely adopted in various applications such as \textit{de novo} drug design \cite{li2021smiles}, molecular property prediction \cite{chithrananda2020chemberta}, and protein-ligand binding affinity prediction \cite{huang2020moltrans}. BPE starts with an initial vocabulary of individual characters and iteratively merges the most frequently occurring pairs until a desired vocabulary size is reached. Another subword tokenization method used in molecular design \cite{blanchard2023adaptive} is WordPiece \cite{wu2016google}, which is similar to BPE in that it also begins with an initial vocabulary and continues merging pairs of tokens. However, unlike BPE, WordPiece selects the pair that maximizes the likelihood of the training data, rather than simply the most frequent one. In contrast to both BPE and WordPiece, Unigram \cite{kudo-2018-subword} is a top-down tokenization method that starts with an initial vocabulary of all possible words and reduces it based on the likelihood of the unigram model, and it has recently been used in molecular fingerprinting \cite{abdel-aty_large-scale_2022}.

We adopt three widely utilized subword tokenization algorithms, namely Byte Pair Encoding (BPE), WordPiece, and Unigram, which have been successfully applied in computational drug discovery tasks \cite{li2021smiles,chithrananda2020chemberta, huang2020moltrans, blanchard2023adaptive, abdel-aty_large-scale_2022} and compare the vocabularies generated by these methods. However, due to the large number of chemical words in the resulting vocabularies, it is impractical to interpret them all manually. Therefore, we propose a novel language-inspired pipeline to identify key chemical words that are strongly associated with protein-ligand binding. In natural languages, each genre or period has a certain distribution of words. Similarly, each protein family can be expected to have a certain distribution of substructures (or chemical words) to which it would bind. Therefore, we focused on the high affinity ligands of protein families to build protein family specific vocabularies and analyzed the chemical significance of subwords for  the protein family. Protein-ligand binding is selected as the chemical investigation perspective since strong binding relationships are fundamental in drug discovery pipelines, are widely available in the literature, and naturally group chemicals by their binding targets. The pipeline processes a protein-ligand binding affinity dataset and identifies ten chemical words for each protein or protein family in the dataset. We find that while the vocabularies generated by different subword tokenization algorithms differ in words, lengths, and validity, the identified key chemical words are similar, as measured by the mean edit distance between protein or family-specific top ranking subwords with different vocabularies. We also observe that the selected words are protein or family-specific, often associated with only one protein/family, and significantly different from the words identified for weak binders.
As a case study, we examine the selected chemical words for a number of important drug target families and find that the top-ranking chemical words are associated with the known pharmacophores and functional groups of the protein families. Notably, for the aldehyde dehydrogenase 1 enzyme family, 
the chemical words selected by the proposed algorithm were found to improve the drug-likeness of the molecules.
Our results corroborate the chemical word-based models in computational drug discovery and are a step toward interpreting the computational drug discovery models from a pharmaceutical chemistry perspective.

\section{Materials and methods}

In this work, we build a pipeline to analyze the chemical significance of the chemical words of high affinity ligands of proteins or protein families. The vocabularies are created for each protein family based on the hypothesis that the important subwords of each protein family will be specific to the protein family. The top ranking words for the ligands of carbonic anhydrases and casein kinase 1 gamma in BindingDB (BDB), as well as pyruvate kinase M2 and aldehyde dehydroganse 1 from Lit-PCBA are examined for chemical relevance.

\subsection{Segmenting Chemicals into Chemical Words}
\label{sec:chemical_word_segmentation}

In this study, we adopt three commonly used subword tokenization algorithms: BPE \cite{sennrich2015neural}, WordPiece \cite{wu2016google}, and Unigram \cite{kudo-2018-subword}, and learn vocabularies with sizes of 8K, 16K, and 32K. 
The vocabularies are identified by applying the tokenization algorithms on the SMILES representations of $\sim$2.3M compounds in ChEMBLv27 \cite{mendez2019chembl}, and then, the vocabularies are used to segment the compounds into their chemical words.  

\subsection{Characterizing Chemical Words}

Subword tokenization algorithms, adopted from the field of natural language processing, have recently been widely used to identify chemical words and represent compounds in computational drug discovery studies. However, the identified vocabularies have not been characterized and a comparison of the different methods has not been performed yet. To investigate the chemical words learned by different tokenization methods, we analyze various statistics, such as word length, validity, vocabulary overlaps across the algorithms, and word similarity to the most similar extended functional group, which is a generalized version of traditional functional group and introduced by \citet{lu2021dataset}.

\subsection{Term Frequency - Inverse Document Frequency}

Term Frequency - Inverse Document Frequency (TF-IDF) \cite{schutze2008introduction} is a document vectorization algorithm originally introduced in the field of information retrieval. TF-IDF represents documents based on the importance of each word in the document's vocabulary. TF-IDF postulates that the importance of a word for a document is proportional to the word's frequency in the document and the inverse number of documents the word appears in the corpus. The TF-IDF vector for a document $D$ is formulated as:

\begin{equation}
\vec{D} = [tf_{w_1,D} * idf_{w_1}, \cdots, tf_{w_V,D} * idf_{w_V}]_{1 \times V} 
\end{equation}

\noindent where $tf_{w_i,D}$ is the count of the $i^\text{th}$ word in the vocabulary, $w_i$, in $D$.
$idf_{w_i}$ is the natural logarithm of the number of documents in the corpus divided by the number of documents in which $w_i$ is present; and $V$ is the number of different words in the corpus, \textit{i.e.,} the vocabulary size. However, the naive TF-IDF weighting may consider twenty occurrences of a word in a document twenty times more significant than a single occurrence, which may not be accurate. To address this issue, a sublinear term frequency scaling variant of the algorithm has been introduced \cite{schutze2008introduction} , which uses the logarithm of the term frequency and assigns a weight given by:

\begin{equation}
tf_s(w, D) = \begin{cases}1+\log  tf_{w,D}  & \text { if } tf_{w,D} >0 \\ 0 & \text { otherwise }\end{cases}
\end{equation}

\noindent where $tf_{w,D}$ is the count of word $w$ in document $D$ and $tf_s(w, D)$ is its scaled counterpart used in the importance computation instead.

\subsection{Datasets}

We used three different datasets of protein-ligand affinity.  

The BDB dataset contains affinity information for 31K interactions of 924 compounds and 490 proteins from 81 protein families. The BDB dataset reports the protein-compound affinities in terms of $pK_d$ 
and has been used in previous protein-compound interaction prediction studies \cite{ozccelik2021chemboost,ozccelik2021debiaseddta}. To identify protein family-specific words, we map each protein in this dataset to its corresponding PFAM families \cite{mistry2021pfam} using the InterPro API \cite{paysan2023interpro}.  

LIT-PCBA \cite{tran2020lit}  includes 15 target proteins and 7844 active and 407,381 inactive compounds. The dataset is specifically curated for virtual screening and machine learning with efforts made to ensure it is unbiased and realistic.

ProtBENCH \cite{atas2023approach} contains protein family-specific bioactivity datasets. These datasets include interactions belonging to different protein superfamilies, including membrane receptors, ion channels, transporters, transcription factors, epigenetic regulators, and enzymes with five subgroups (i.e., transferases, proteases, hydrolases, oxidoreductases, and other enzymes). The family subsets have varying sizes in terms of interactions ( from 19K to 220K), number of proteins ( from 100 to 1K), and number of compounds (from 10K to 120K).

\subsection{Identifying the Key Chemical Words for Strong Protein-Ligand Binding}

\begin{figure}[H]
    \centering
    \includegraphics[width=\textwidth]{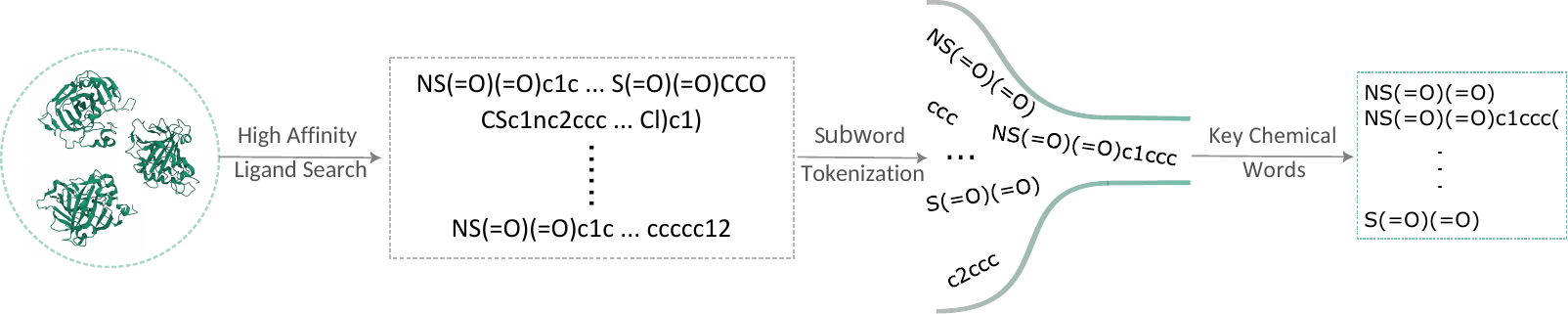}
   
    \caption{\textit{The proposed pipeline to identify key chemical words for high affinity to a protein family.} For each protein family, the pipeline first extracts the SMILES representations of compounds that bind to the protein family with high affinity. Next, the SMILES strings are segmented into their chemical words via the tokenization algorithms such as BPE, and each high-affinity compound list is modeled as a document composed of SMILES "sentences". Last, the compound documents are vectorized via TF-IDF, which assigns an importance score to each chemical word per protein family, and the ten chemical words with the highest TF-IDF scores are identified as key chemical words. Key chemical words for carbonic anhydrase, casein kinase 1 gamma,  pyruvate kinase M2, and aldehyde dehydrogenase 1 enzyme systems are analyzed further to interpret their chemical significance.}
    \label{fig:pipeline}
\end{figure}

Here we propose a novel pipeline to identify key chemical words for strong binding to both individual proteins and to protein families. The proposed algorithm is language-inspired and postulates, similar to TF-IDF, that if a chemical word is common and unique to the strong binders of a protein or protein family, then it signifies a key chemical substructure for binding. 
Accordingly, the algorithm first identifies strong binders for each protein or protein family. While LIT-PCBA already distinguishes the interactions as strong and weak binders, strong binders in other datasets are identified using predefined thresholds such as a $pK_d$ value higher than 7 for BDB and a bioactivity score (pChEMBL) greater than the median score for ProtBENCH. 
Next, strong binders of each protein or protein family are represented as a document, in which each strong binding compound is a sentence composed of chemical words identified via the algorithms described in the Segmenting Chemicals into the Chemical Words section.
Finally, documents containing the SMILES representations of the high affinity ligands of each protein or protein family are vectorized with TF-IDF and the chemical words are ranked based on their TF-IDF scores. The proposed pipeline is illustrated in Figure ~\ref{fig:pipeline}.

\subsection{Comparing Key Chemical Words}
While the proposed pipeline aims at identifying key chemical words for strong binders of individual proteins or protein families, it can also be used to compute word importance scores for weak binders. To compare the importance scores of weak binders with those of strong binders, we apply the pipeline on weak binders by considering weak interactions as documents associated with proteins or protein families. Next, we conduct Wilcoxon rank-sum tests (p < 0.05) on a target level to compare the importance scores of chemical words associated with strong binding to each protein or family with those associated with weak binding to that particular target.

To investigate whether the chemical words identified by using different vocabularies are similar, we compute mean edit distance scores between selected chemical words using the following equation:

\begin{equation}
\frac{1}{|A_p|} \sum_{h_{A_p} \in A_p} \min_{h_{B_p} \in B_p} d\left(h_{A_p}, h_{B_p}\right)
\end{equation}

\noindent where $A$ and $B$ denote chemical word vocabularies, with $A_p$ and $B_p$ being the subsets identified for a specific protein p. The individual words in these subsets are represented by $h_{A_p}$ and $h_{B_p}$. The Levenstein edit distance function, $d(h_{A_p}, h_{B_p})$, is used to determine the distance between two words. This score allows us to compare the similarity of chemical words identified for the same protein or protein family across different vocabularies.

\section{Results}

In this study, we investigated the characteristics of the chemical vocabularies generated by different subword tokenization algorithms (Unigram, BPE, and WordPiece). We also conducted case studies on protein families to analyze the impact of key chemical words on binding.

\subsection{Characterizing Vocabularies}

While subword tokenization algorithms are commonly used for identifying chemical words, it is important to investigate whether they yield consistent results across different algorithms. In this study, we compared the vocabularies of three commonly used subword tokenization algorithms: Unigram, BPE, and WordPiece.

\paragraph{Chemical words are shorter than conventional functional groups}

\begin{figure}[ht]
    \centering
    \includegraphics[width=0.5\textwidth]{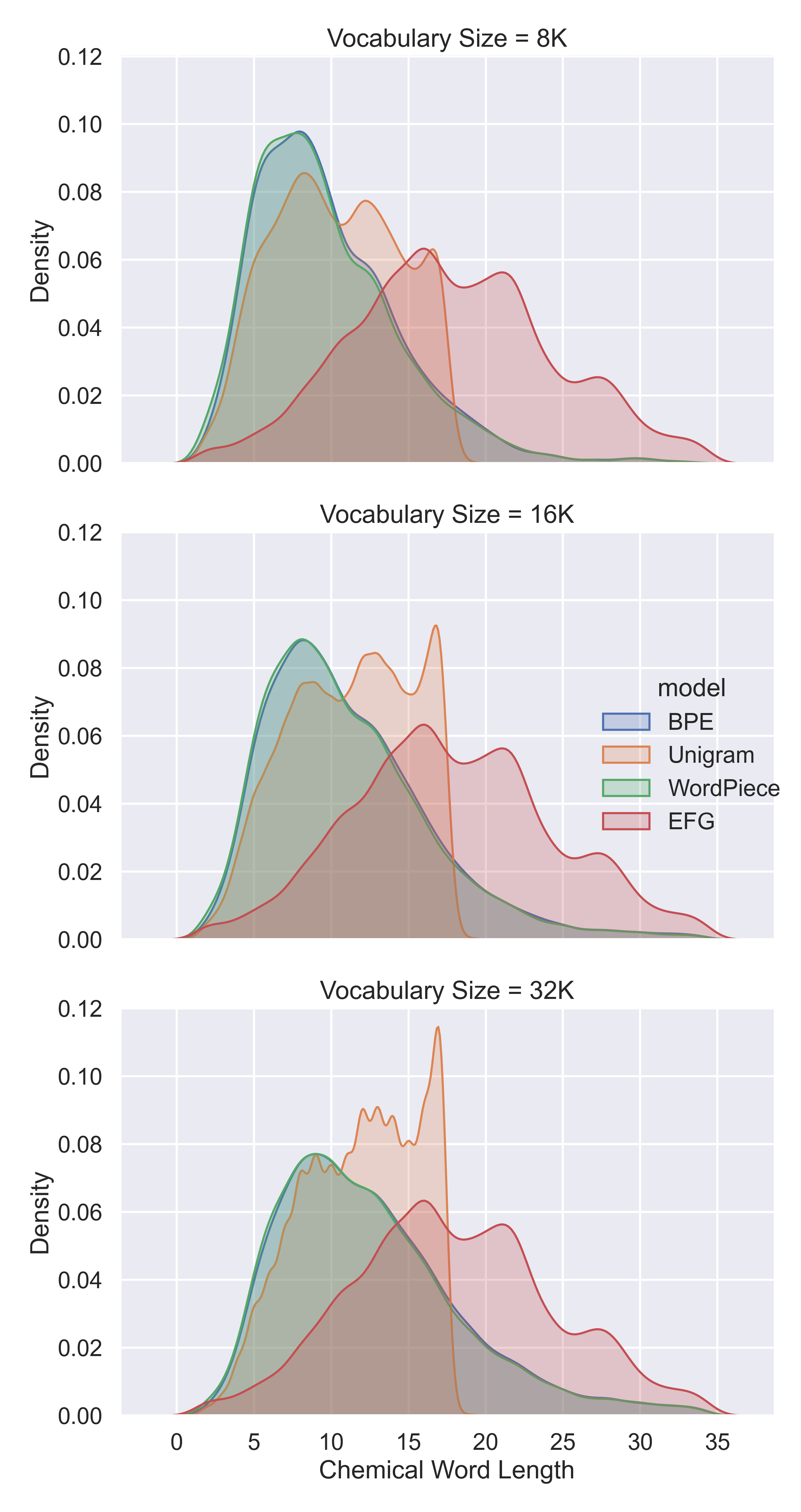}
    \caption{Length distribution of words identified with different subword tokenization algorithms and varying vocabulary sizes. }
    \label{fig:length}
\end{figure}

The mean length of chemical words varied from 9 to 13, depending on the subword tokenization algorithm and vocabulary size. Words identified using Unigram are longer, with the longest word having a length of 19. Extended functional groups (EFG), on average, are even longer with a mean length of approximately 19. There is a trend of increase in mean length illustrated in Figure \ref{fig:length}, where the distribution of word lengths across the tokenization methods shows a slight shift towards longer words with increasing vocabulary size.

\paragraph{BPE and WordPiece yield vocabularies with a larger number of valid words compared to Unigram} 

\begin{wrapfigure}{r}{0.5\textwidth}
  \centering
    \includegraphics[width=0.5\textwidth]{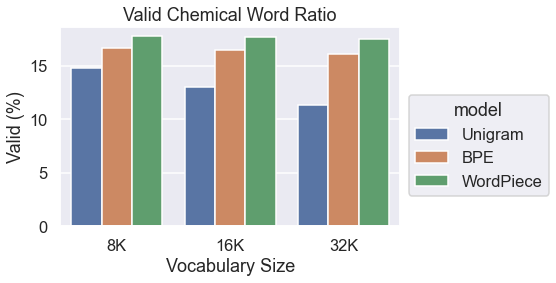}
    \caption{Valid word ratio of vocabularies identified with the subword tokenization algorithms, Unigram (blue), BPE (orange), and Wordpiece (green) and vocabulary sizes of 8K, 16K and 32K.}
    \label{fig:validity}
\end{wrapfigure}

Although subword tokenization algorithms can identify chemical words, it is important to note that not all identified words may be chemically valid. 
However, having valid words can be desirable as it may enhance the interpretability of models and facilitate approaches such as scaffold hopping.
We used RDKit \cite{landrum2006rdkit} to assess validity, which considers a molecule valid if it complies with the principles of chemical bonding and doesn't exhibit any unusual or unfeasible structures. 

The valid word ratio of the compared vocabularies ranges from 11\% to 17\%, with Unigram vocabularies having the lowest ratio, and WordPiece vocabularies achieving the highest ratio as shown in Figure~\ref{fig:validity}. While BPE and WordPiece vocabularies achieve a comparable valid ratio which is also consistent across varying vocabulary sizes, the valid ratio of Unigram vocabularies decreases with increasing vocabulary size.

\paragraph{Do the subword tokenization algorithms yield the same words?} 

\begin{wrapfigure}{r}{0.5\textwidth}
\centering
\includegraphics[width=0.3\textwidth]{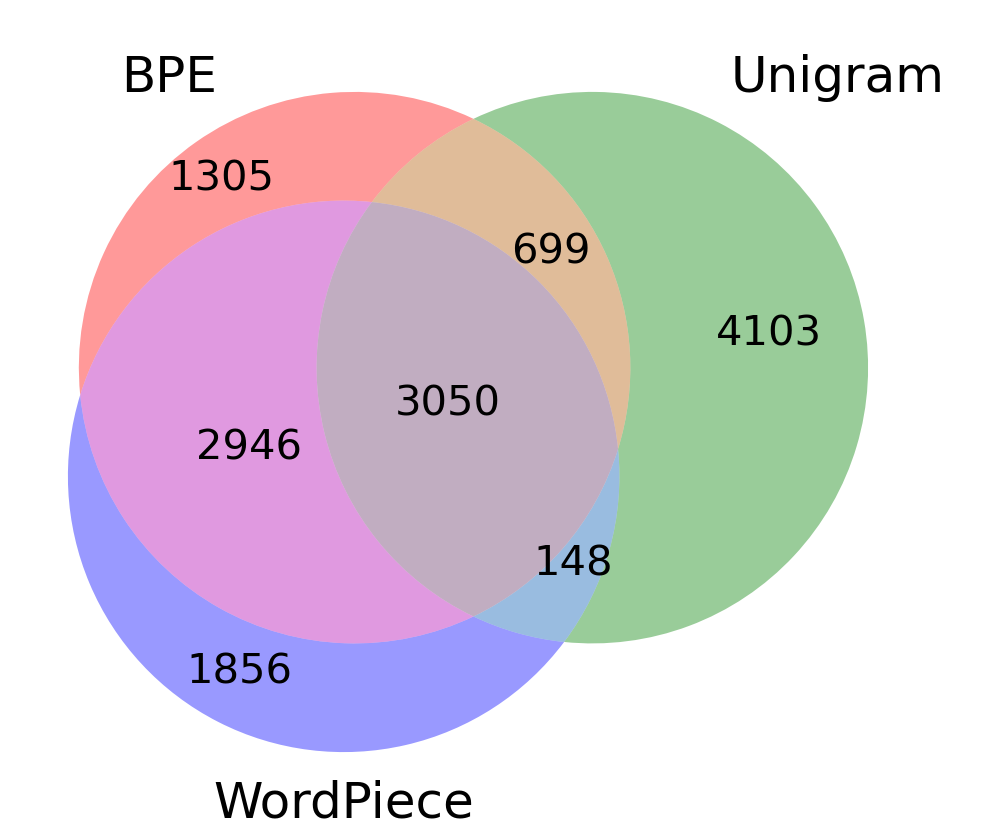}
\caption{Common chemical words across 8K-sized vocabularies.}
\label{fig:intersection}
\end{wrapfigure}

Subword tokenization algorithms, despite having similar approaches, can result in different vocabularies. Figure~\ref{fig:intersection} shows the number of common words between 8K-sized vocabularies produced by different tokenization methods.

\begin{figure}[ht]
\centering
\includegraphics[width=0.4\linewidth]{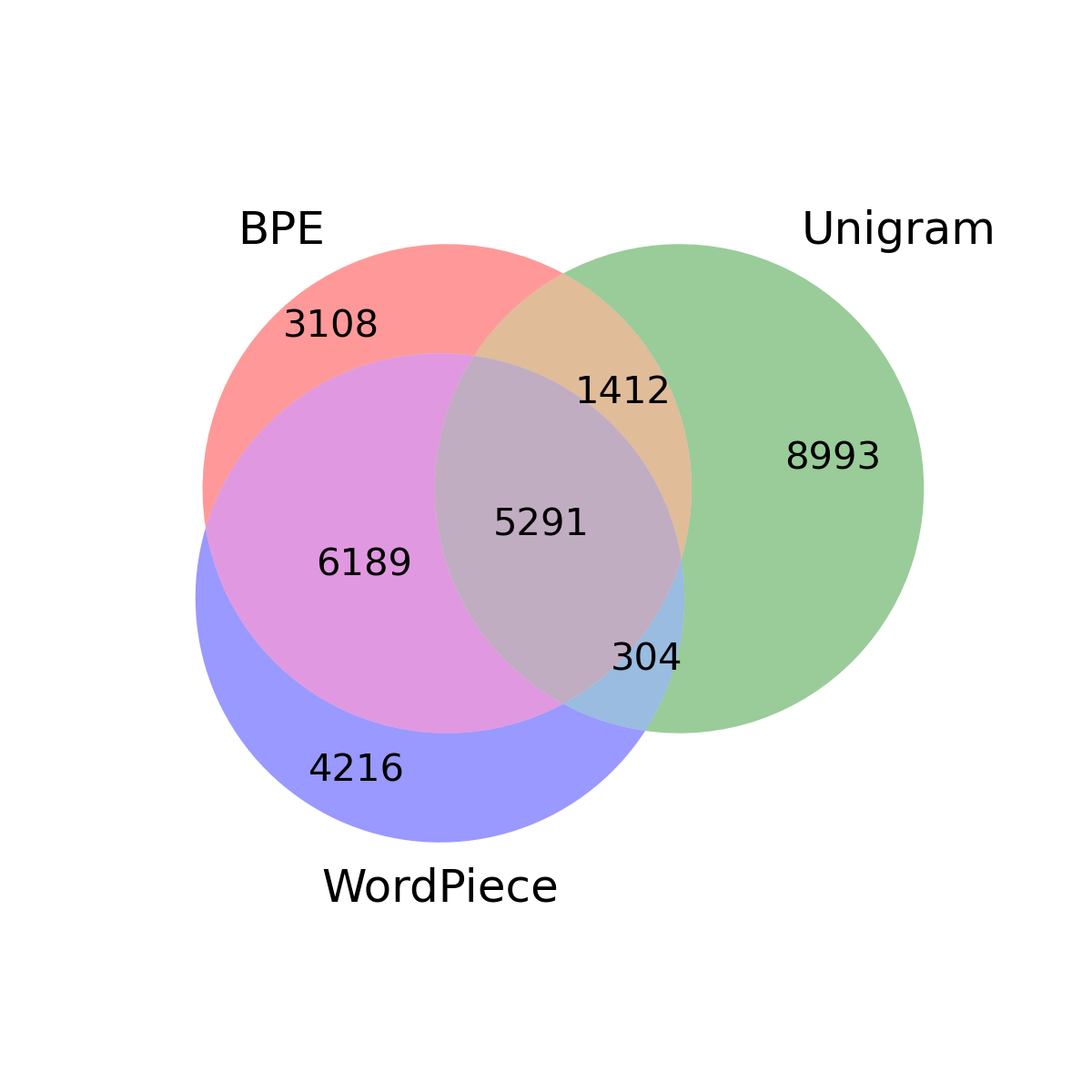}
\caption{Common chemical words across 16K-sized vocabularies.}
\label{fig:intersection16K}
\end{figure}

\begin{figure}[ht]
\centering
\includegraphics[width=0.4\linewidth]{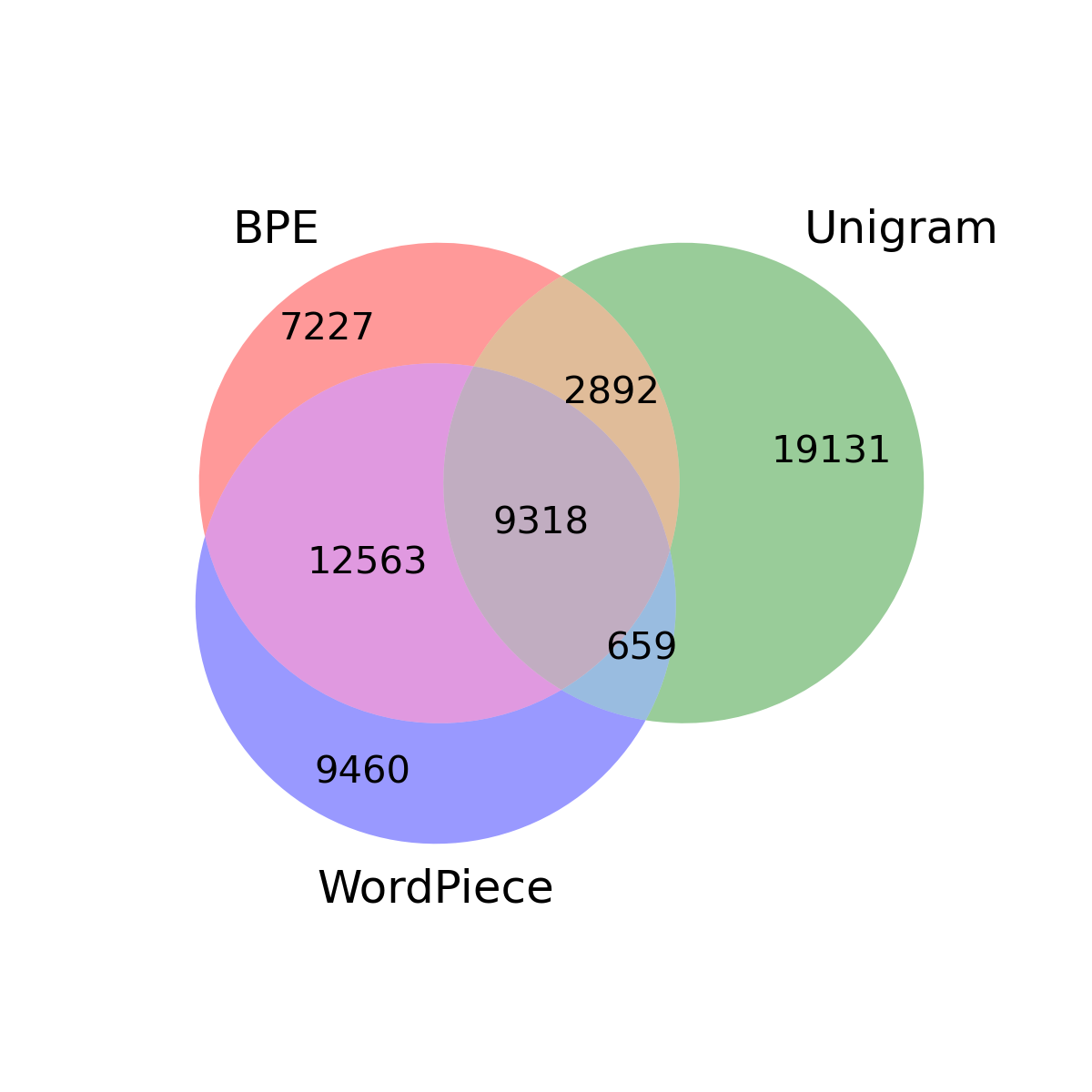}
\caption{Common chemical words across 32K-sized vocabularies.}
\label{fig:intersection32K}
\end{figure}

BPE and WordPiece vocabularies share approximately 80\% of their words, while Unigram vocabularies share at most about 47\% of their words with other tokenization algorithms. The ratio of common words is 38\% in 8K-sized vocabularies, but it decreases to 33\% in 16K and 29\% in 32K sized vocabularies (see Figure~\ref{fig:intersection16K} and ~\ref{fig:intersection32K} for 16K and 32K sized vocabularies).

\hfill \break

\subsection{Identifying Chemical Key Words}

\begin{wrapfigure}{r}{0.5\textwidth}
    \centering
    \includegraphics[width=0.5\textwidth]{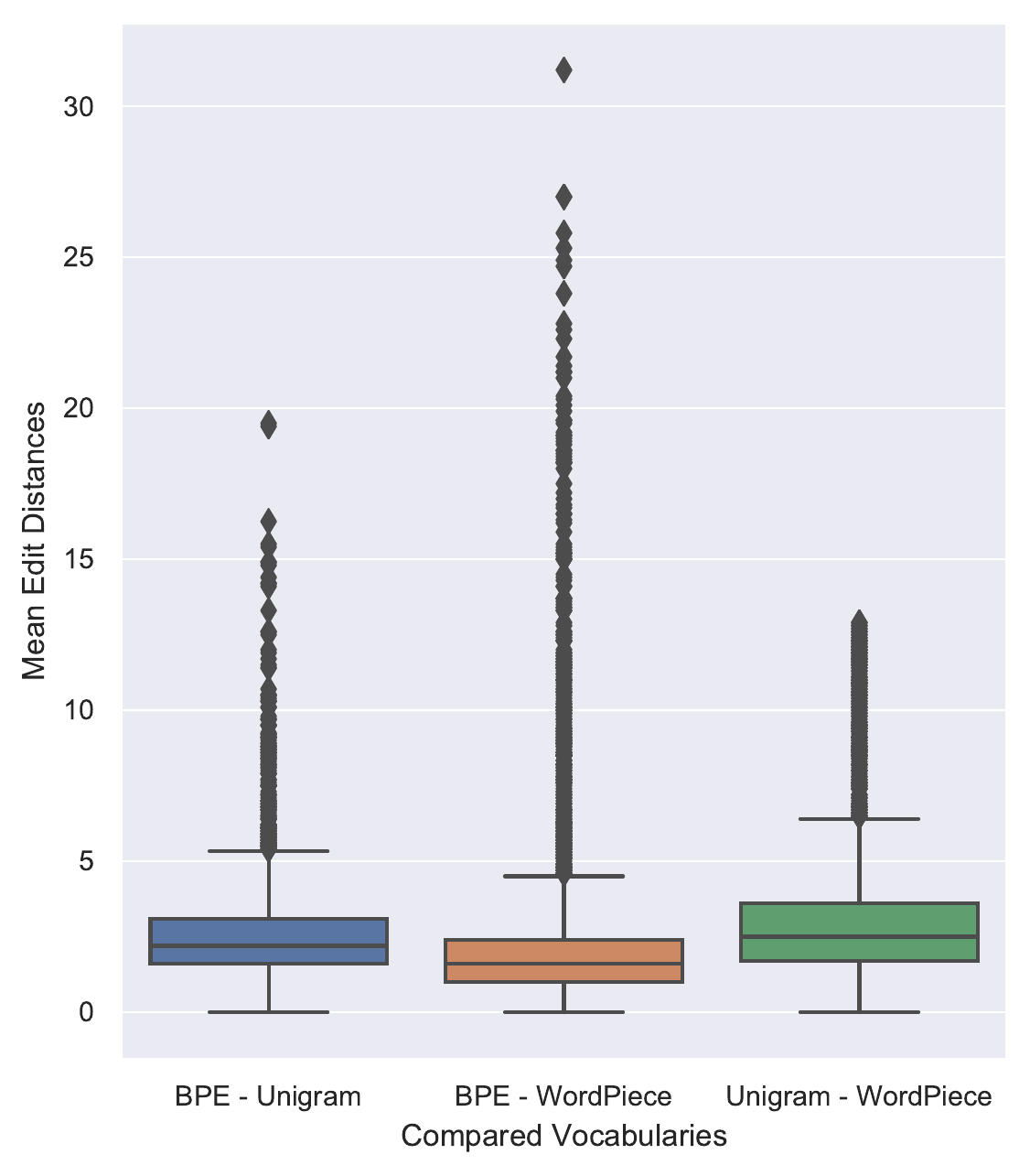}
    \caption{Distribution of mean edit distances between sets of chemical key words identified using 8K-sized vocabularies}
    \label{fig:vocab_comparison}
\end{wrapfigure}

Our proposed pipeline was applied to identify chemical key words for protein families in the BDB dataset, as well as proteins in the LitPCBA and ProtBENCH datasets. To identify chemical key words, we ran the pipeline on each dataset using the three different tokenization algorithms with three distinct vocabulary sizes. We then evaluated the similarity of the identified key chemical words across different vocabularies by calculating the mean edit distances between these words for each target protein or protein family. Figure~\ref{fig:vocab_comparison} shows the distribution of the mean edit distance between 8K-sized vocabularies. Although the words selected by different vocabularies were not identical, they were quite similar, as indicated by the mean edit distance scores primarily ranging between 0 and 6, compared to a mean edit distance score of 6 for sets of words randomly selected from distinct vocabularies. For the majority of the targets, the mean edit distance was quite low, ranging from 2 to 4, depending on the compared vocabularies. This finding indicates that the three different tokenization algorithms using different sized vocabularies yielded similar chemical key words. Consequently, we used the BPE vocabulary with 8K words for the remainder of our analysis.

\begin{figure}[!ht]
    \centering
    \includegraphics[width=0.6\textwidth]{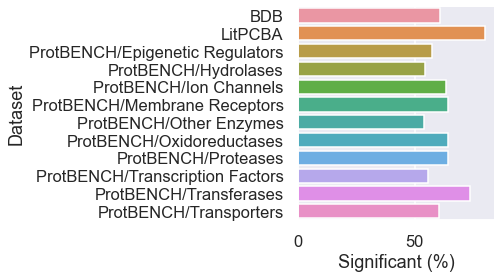}
    \caption{Percentage of families/proteins with significantly different chemical key words across datasets.}
    \label{fig:significance}
\end{figure}

The proposed method identified key words ranging from $\sim$150 to  $\sim$ 3000, depending on the number of proteins or protein families in the dataset (see Table~\ref{tab:highlights}). The mean number of associations per identified chemical key word varies between $\sim$1 and $\sim$3, indicating that these words are indeed protein or family specific. However, due to the large vocabulary sizes, it was not feasible for us to manually assess all the chemical key words. Therefore, we compared the importance scores of these words between strong and weak binders for each protein or family.  Our analysis revealed that these scores were significantly different for at least half of the families or proteins across datasets, as shown in Figure~\ref{fig:significance}.

\begin{table}[!ht]
\centering
\caption{Number of identified chemical key words and mean number of associated protein/family per word across datasets.}
\label{tab:highlights}
\begin{tabular}{@{}lcc@{}}
\toprule \textbf{Dataset}                & \textbf{\# Chemical Key Words} & \textbf{Mean Associations} \\ \midrule
BDB                             & 531                  &  3.01                                      $\pm$ 2.41         \\
LitPCBA                         & 147                   & 1.02                                      $\pm$ 0.14         \\
ProtBENCH/Epigenetic Regulators & 538                   & 1.53                                      $\pm$ 1.10         \\
ProtBENCH/Hydrolases            & 2191                  & 1.94                                      $\pm$ 1.91         \\
ProtBENCH/Ion Channels          & 1068                  & 1.45                                      $\pm$ 0.99         \\
ProtBENCH/Membrane Receptors    & 2912                  & 1.80                                      $\pm$ 1.28         \\
ProtBENCH/Other Enzymes         & 1430                  & 1.59                                      $\pm$ 1.27         \\
ProtBENCH/Oxidoreductases       & 1809                  & 1.59                                      $\pm$ 1.12         \\
ProtBENCH/Proteases             & 1545                  & 1.82                                      $\pm$ 1.57         \\
ProtBENCH/Transcription Factors & 664                   & 1.41                                      $\pm$ 0.93         \\
ProtBENCH/Transferases          & 2969                  & 2.25                                      $\pm$ 3.33         \\
ProtBENCH/Transporters          & 865                   & 1.33                                      $\pm$ 0.70         \\ \bottomrule
\end{tabular}
\end{table}

\subsection{Case studies}
We selected two protein families (carbonic anhydrases and casein kinase 1 gamma) from BDB dataset and two protein families (pyruvate kinase M2 and aldehyde dehydrogenase 1) from Lit-PCBA dataset as case studies and analyzed the chemical properties of the key chemical words and studied their impact on binding.

\paragraph{Carbonic Anhydrases}

Carbonic anhydrase (CA) is a metalloenzyme found in both eukaryotes and prokaryotes and is involved in many metabolic activities such as ion transport, pH balance, and gluconeogenesis \cite{supuran2003carbonic}. Since the discovery that CA enzymes are involved in important processes in the organism, they have become therapeutic targets for drug development. CA inhibitors are used clinically for diuretic, antiglaucoma, antiepileptic and antiobesity effects. New compounds with many anticancer effects that selectively inhibit CA isoenzymes have been designed and developed \cite{supuran2010carbonic}. 

The algorithm proposed the following chemical words as CA pharmacophores or functional groups, in decreasing order of TF-IDF score: \texttt{c1c(F)c(F)}, \texttt{c(F)c1F}, \texttt{S(N)(=O)=O}, \texttt{CC(=O)c1ccc(}, \texttt{c(S}, \texttt{NS(=O)(=O)c1ccc(},\texttt{NS(=O)(=O)}, \texttt{c(Cl)c1)}, \texttt{c1nc2ccccc2n1}, and \texttt{CSc1n}.

Here, the words \texttt{S(N)(=O)=O},  \texttt{NS(=O)(=O)}, \texttt{NS(=O)(=O)c1ccc(}, and \texttt {c1nc2ccccc2n1}  are either components of chemical moieties or are already complete chemical moieties. These chemical words are recognized as sulfonamide, sulfonamide, aryl substituted sulfonamide, and 1\textit{H}-benzo[\textit{d}]imidazole.  We noted that the chemical words \texttt{NS(=O)(=O)} and \texttt{S(N)(=O)=O} are both sulfonamides, although their SMILES representations are different. 

\begin{figure}[H]
\centering
\begin{subfigure} 
\centering
\includegraphics[width=0.8\linewidth]{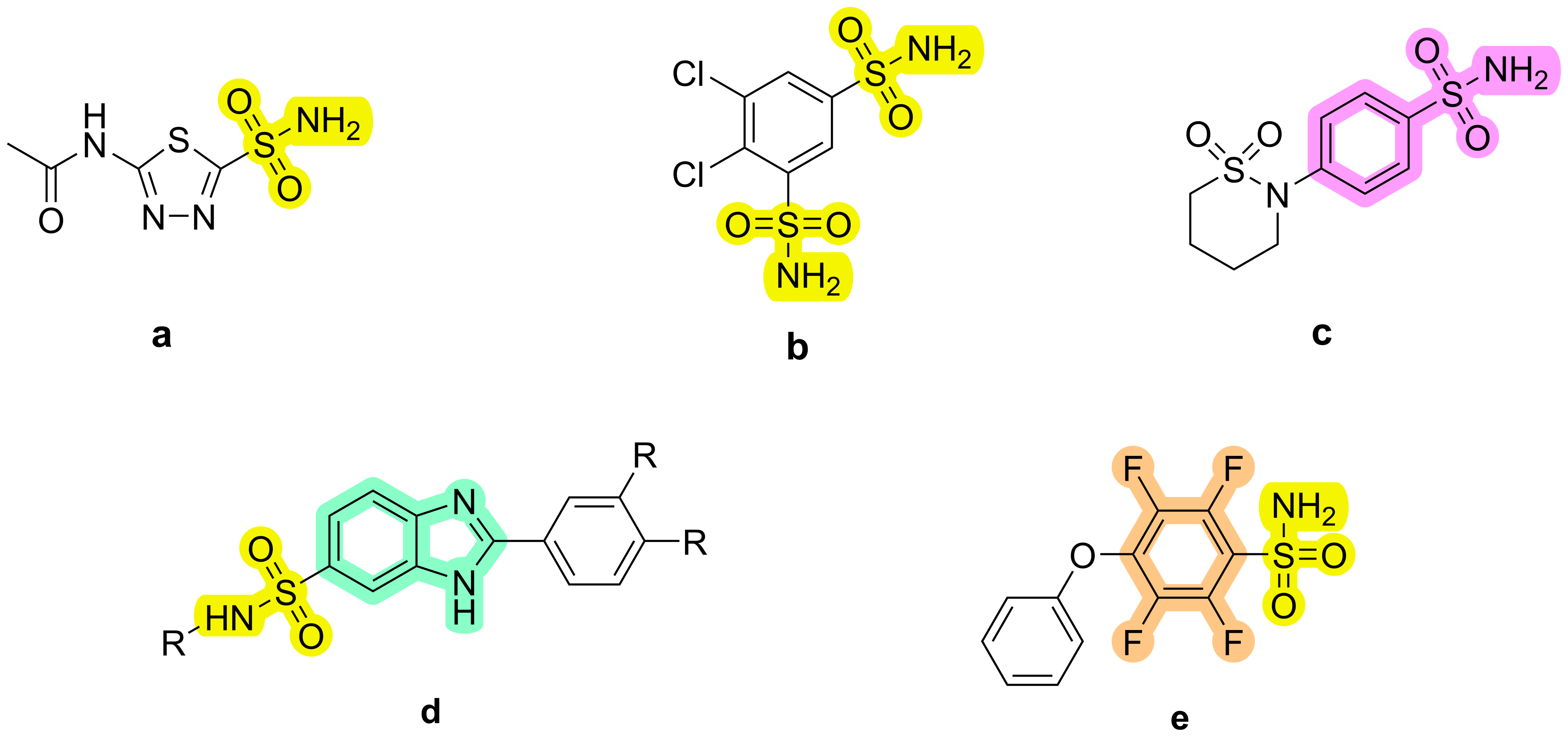}
\end{subfigure}
\caption{\textit{Interpreting the key chemical words for carbonic anhydhrases.} 2D representations of \textit{acetazolamide} (\textbf{a}), \textit{dichlorphenamide} (\textbf{b}), \textit{sultiame} (\textbf{c}),
\textit{2-subtituated-benzimidazole-6-sulfonamide derivative} (\textbf{d}), and
\textit{2,3,5,6-tetrafluoro-4-phenoxybenzenesulfonamide} (\textbf{e}). 
The chemical words, \texttt{S(N)(=O)=O}, \texttt{NS(=O)(=O)c1ccc}, \texttt{c1nc2ccccc2n1},  and \texttt{c1c(F)c(F)} proposed by the algorithm as the key chemical words (sulfonamide, aryl-substituted sulfonamide, 1\textit{H} -benzo[\textit{d}]imidazole, and fluorine substituted aromatic ring), are marked in yellow, pink, green and orange on the compounds, respectively.
}
\label{fig:ca_words}
\end{figure}

The sulfonamide (\texttt{RSO2NH2}) group interacts with the Zn(II)  ion in the active site of the CA enzyme and constitutes the main pharmacophoric group for CA inhibition. Sulfonamide derivatives bearing aromatic or heteroaromatic rings are potent inhibitors of the CA enzyme \cite{casini2003carbonic, supuran2010carbonic}. Systemic inhibitors such as acetazolamide (\textbf{a}) and dichlorophenamide (\textbf{b}) have long been used clinically as antiglaucoma agents\cite{supuran2003carbonic}. Sulthiame (\textbf{c}) is a clinically used antiepileptic agent that shows inhibition of many CA isoenzymes in the brain \cite{sugrue2000pharmacological}. The 2D structures of the clinically used CA enzyme inhibitors acetazolamide (\textbf{a}), dichlorophenamide (\textbf{b}) and sulthiame (\textbf{c}) are shown in Figure \ref{fig:ca_words}. Isoform selectivity of CA inhibitors is important to minimize side effects and increase therapeutic efficacy. For this purpose, researchers have designed a new generation of drugs that provide isoform selectivity. For example, CA II is an important target in the treatment of glaucoma, while CA IX and XII isoenzymes are mostly found in tumor cells. Second generation antiglaucoma drugs such as dorzolamide and brinzolamide are potent selective CA II inhibitors \cite{fares2020discovery}. In addition, studies have shown that benzenesulfonamide derivatives show strong inhibition of highly tumor-specific CA IX and XII isoenzymes \cite{supuran2013carbonic}. These compounds contain sulfonamide and aryl-substituted sulfonamide structures, which were also identified by the algorithm as key chemical words for the CA enzyme family.

The key chemical word \texttt{c1nc2ccccc2n1} found by the algorithm is notable among CA inhibitors with promising selectivity profiles. This is because CA enzyme inhibitors containing the benzimidazole structure represented by the word have been shown to be derivatives with high potential in the development of novel anti-cancer agents \cite{milite2019novel,mulugeta2022synthesis}. Figure \ref{fig:ca_words} shows the 2D structure of a series of newly synthesized 2-subtituted-benzimidazole-6-sulfonamide (\textbf{d}) derived compounds with high potential against hCA IX and/or XII isoforms \cite{milite2019novel}.

The words \texttt{c1c(F)c(F)}, \texttt{c(F)c1F}, \texttt{c(Cl)c1)} do not correspond to a full chemical moiety since ring closures are missing, but they are part of fluorine and chlorine substituted aromatic ring structures. In the literature, studies have investigated the thermodynamics, CA inhibitory effects, and binding affinities of benzenesulfonamide structures bearing fluorine or chlorine \cite{scott2009thermodynamic,kim2000contribution,slawinski2014carbonic}. A study by \citet{dudutiene20134} showed that all fluorinated benzenesulfonamide compounds had nanomolar binding potency against the tested CA enzymes. Furthermore, fluorinated benzenesulfonamides were found to have higher binding potency than non-fluorinated compounds. Figure \ref{fig:ca_words} shows the 2D structure of 2,3,5,6-tetrafluoro-4-phenoxybenzenesulfonamide (\textbf{e}), a polyfluorinated benzenesulfonamide derivative.

Similarly, \texttt{CC(=O)c1ccc(}, which is not fully determined due to missing ring closures and is not a meaningful word for the CA family. Last \texttt{c(S} and \texttt{CSc1n} do not designate any chemical substructure.

\paragraph{Casein Kinase 1 gamma}
\begin{wrapfigure}{r}{0.5\textwidth}
  \begin{center}
    \includegraphics[width=0.35\textwidth]{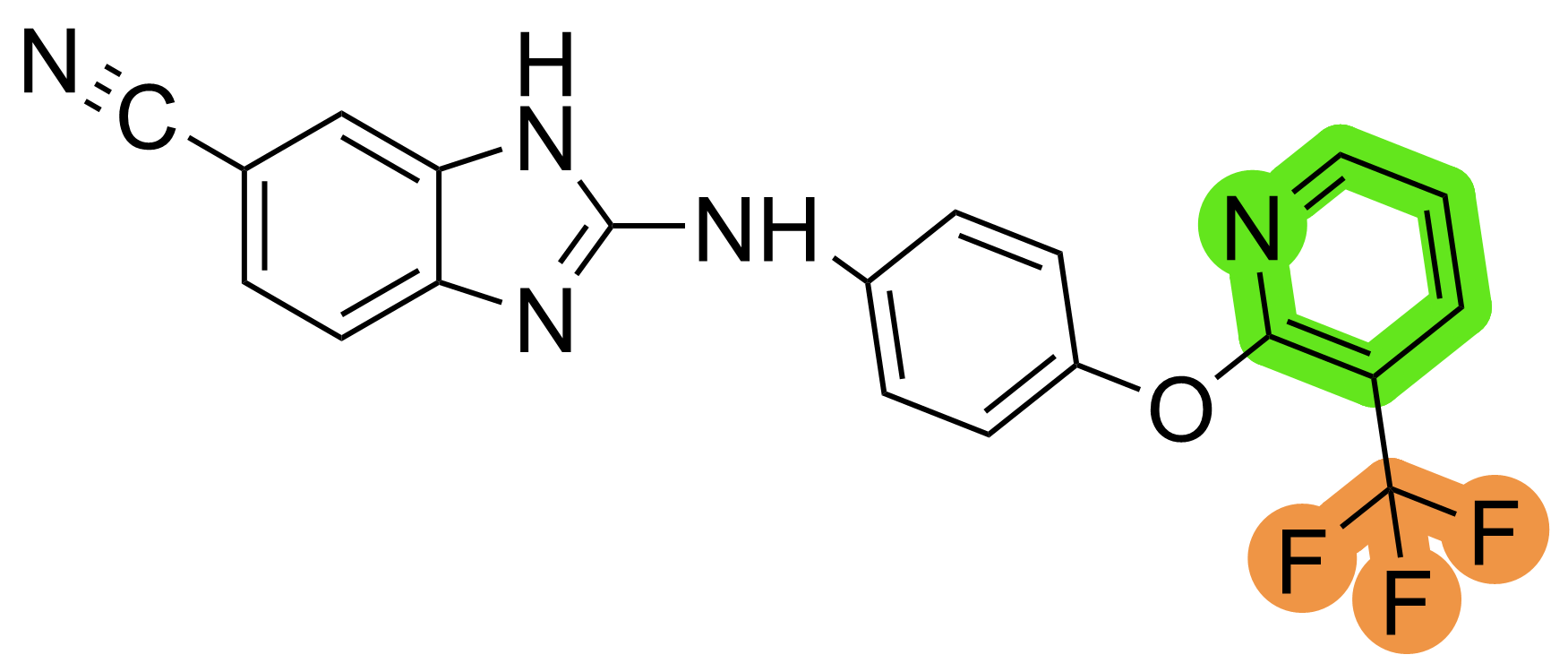}
  \end{center}
  \caption{\textit{Interpreting chemical words of caseine kinase 1 gamma enzyme system.}
A 2-phenylamino-6-cyano-1\textit{H}-benzimidazole derivative bearing trifluoromethyl at the pyridine ring is a potent and selective CK1 gamma inhibitor. The chemical words \texttt{c1ccncc1} and \texttt{C(F)(F)F)CC1} proposed by the algorithm correspond to pyridine (green) and trifluoromethyl (orange) groups.  }
  \label{fig:casein_words}
\end{wrapfigure}
Casein kinase 1 (CK1) isoforms are important therapeutic targets for diseases such as Alzheimer's disease (AD), amyotrophic lateral sclerosis (ALS), and cancer  \cite{gunby2007oncogenic,janovska2020targeting,li2021recent}.
Casein kinase 1 gamma (CK1$\mathcal{\gamma}$), one of the isoforms of CK1, plays a role in biological processes such as signal transduction, DNA repair, cell division and circadian rhythm in cells \cite{zhai1995casein}.  CK1$\mathcal{\gamma}$ domain family refers to a domain found in the structure of proteins and associated with CK1$\mathcal{\gamma}$. The C-terminal region may play a role in the interaction of the protein with target molecules. Drug studies focus on developing compounds or inhibitors that can bind to this region by targeting CK1$\mathcal{\gamma}$ \cite{venerando2010isoform}.  

Designing isoform-specific small molecules for members of the CK1 family is important for therapeutic strategies in human cancer. For this purpose, 2-phenylamino-6-cyano 1\textit{H}-benzimidazole derivatives were synthesized as isoform-selective CK1$\mathcal{\gamma}$ inhibitors by \citet{hua20122}. Activity assays showed that the compound carrying a trifluoromethyl group on the pyridine ring shown in Figure~\ref{fig:casein_words} is the most potent CK1$\mathcal{\gamma}$ inhibitor. More importantly, the compound demonstrated high selectivity over other CK1 isoforms such as CK1 alpha and delta. The pyridyl group is involved in an "edge-to-face" hydrophobic interaction with Pro331 \cite{hua20122}.

\begin{figure}[ht]
\centering
\begin{subfigure} 
\centering
\includegraphics[width=0.8\linewidth] {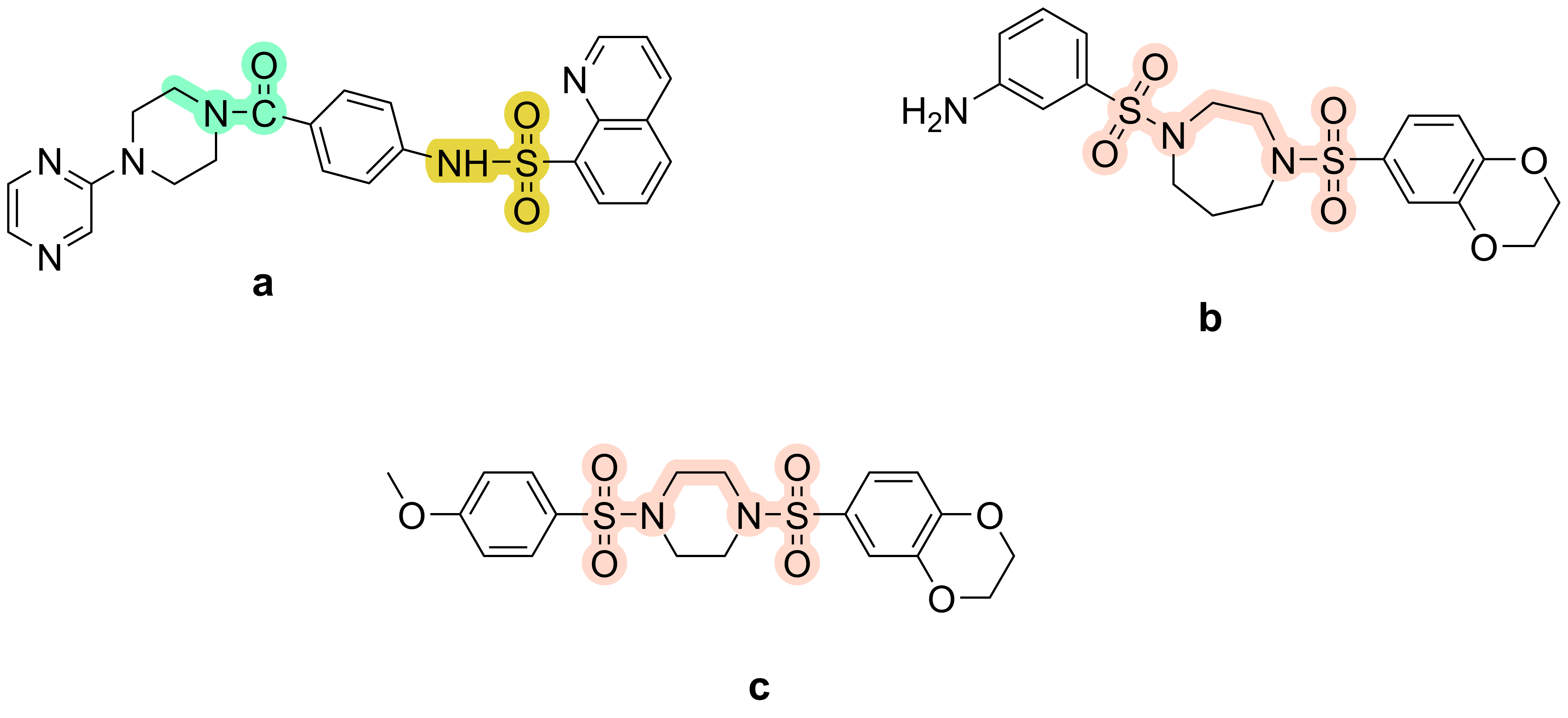}
\end{subfigure}
\caption{\textit{Interpreting chemical words of pyruvate kinase M2 enzyme system.} N-(4-{[4-(pyrazin-2-yl)piperazin-1-yl]carbonyl}phenyl)quinoline-8-sulfonamide (\textbf{a}), 3-{[4-(2,3-dihydro-1,4-benzodioxin-6-ylsulfonyl)-1,4-diazepan-1-yl]sulfonyl}aniline (\textbf{b}), and 1-(2,3-dihydro-1,4-benzodioxin-6-ylsulfonyl)-4-[(4-methoxyphenyl)sulfonyl]piperazine(\textbf{c}). The chemical words, \texttt{S(=O)(=O)N2CCN(} (salmon), \texttt{NS(=O)(=O) / S(=O)(=O)N } (dark yellow), and \texttt{C(=O)N(C)} (green) proposed by the algorithm as the key chemical, are marked on the molecules.}
\label{fig:pkm_words}
\end{figure}

\pagebreak
\paragraph{Pyruvate Kinase M2}
Pyruvate Kinase M2 (PKM2) is an enzyme that catalyses pyruvate production and is involved in energy metabolism and biosynthesis. It also functions as a protein kinase that contributes to tumour formation. It is overexpressed in cancerous cells and pyruvate kinase activity has been proposed as a useful approach for disrupting the required metabolic pathways for cancer cell proliferation \cite{dong2016pkm2}. 
The algorithm indicated the following as key chemical words (in decreasing order of TF-IDF score): \texttt{ccc12), S(=O)(=O)N2CCN(, C2=NN(, S(=O)(=O)N, C(=O)N(C), c2nnnn2, SC, c(SCC(=O)N, NS(=O)(=O)}, and \texttt{c5c4}. \texttt{S(=O)(=O)N} and \texttt{NS(=O)(=O)} correspond to the sulfonamide group, which is present in 159 molecules out of the 546 PKM2 inhibitors in the Lit-PCBA dataset. In addition, \texttt{S(=O)}\texttt{(=O)N2CCN(} is a heterocyclic molecule containing an N atom to which the sulfonamide group is linked and \texttt{C(=O)N(C)} represents the amide group. Figure~\ref{fig:pkm_words} shows 
several compounds that have been documented in the literature to target the PKM2 enzyme. The PDB codes of these molecules are N-(4-{[4-(pyrazin-2-yl)piperazin-1-yl]carbonyl}phenyl)quinoline-8-sulfonamide (\textbf{a}) \cite{kung2012small}, 3-{[4-(2,3-dihydro-1,4-benzodioxin-6-ylsulfonyl)-1,4-diazepan-1-yl]sulfonyl}aniline (\textbf{b}) \cite{anastasiou2012pyruvate}, 1-(2,3-dihydro-1,4-benzodioxin-6-ylsulfonyl)-4-[(4-methoxyphenyl)sulfonyl]piperazine (\textbf{c}) \cite{boxer2010evaluation} respectively. One of the oxygen atoms of the sulfonamide group makes a H-bond with the backbone oxygen of Tyr390. Moreover, the oxygen of the amide moiety forms an H-bond with side-chain nitrogen of Lys311 \cite{kung2012small, delabarre2014action}.

\paragraph{Aldehyde Dehydrogenase 1}

Human aldehyde dehydrogenases (ALDHs) are a family of 19 enzymes that oxidise endogenous and exogenous aldehydes to less reactive, more soluble carboxylic acids in a nicotinamide adenine dinucleotide phosphate (NAD(P))+-dependent reaction \cite{januchowski2013role}. ALDH family is involved in various biological processes required for cell survival as well as cell protection such as detoxification of toxic aldehydes, protection from oxidative stress, and regulation of retinoic acids metabolism \cite{singh2013aldehyde}. Overexpression of ALDH isoforms is associated with disease progression and the development of resistance in many cancers such as breast and lung cancer \cite{dinavahi2019aldehyde}. It has been confirmed that ALDHs overexpressed in cancer stem cells (CSCs) promote tumour growth, metastasis, therapeutic resistance, and immune escape.  Therefore, ALDHs stand out as biomarkers for CSCs in several cancers. Consequently, targeting these enzymes may represent a new potential strategy to overcome therapeutic resistance in cancer patients.  The ALDH1 isoenzyme consists of 6 members, namely ALDH 1A1, 1A2, 1A3, 1B1, 1L1, and 1L2. The isoenzymes in this family are used as potential diagnostic markers in various cancers as well as in non-tumour diseases such as alzheimer, parkinson, cataract formation, and diabetes \cite{duan2023aldefluor}. 

In our literature research on the ALDH1 enzyme family, NCT-501 molecule,  which shows  \textit{in vitro} activity against cisplatin-resistant Cal-27 as well as head and neck squamous cell carcinoma cell lines and reduced tumor growth \textit{in vivo}, attracted our attention \cite{yang2015discovery}.  However, its bioavailability is limited due to hepatic metabolism \cite{kulsum2017cancer}. To improve the bioavailability of NCT-501 molecule, we analyzed the chemical words suggested by the algorithm. The \texttt{NC(=S)N} chemical word, which represents the thiourea group, caught our attention due to its known anti-cancer activity \cite{eissa2016design,asati2021pyrazolopyrimidines,shakeel2016thiourea}. After replacing the urea group with the thiourea group, we found that the oral absorption percentage of the molecule increased from 72\% to 82\%. Additionally, the number of hydrogen-bond acceptors decreased from 10 to 9.5, which improved the physicochemical parameters of the molecule in terms of Lipinski's rule of 5
\cite{lipinski2012experimental} (Table~\ref{tab:nct501}). CM026 is a competitive ALDH1A1 inhibitor. Its ineffectiveness when used alone in GROV1 and A2780DK ovarian cancer cell lines is probably due to a low partition coefficient (logP) leading to low cell permeability \cite{morgan2015characterization}. The replacement of the urea group with the thiourea group in CM026 increased the oral absorption percentage from 79.551\% to 90.170\% and increased the logP value from 2.134 to 2.954 (Table~\ref{tab:cm026}).

\begin{table}[ht]
\centering
\caption{Physicochemical properties of NCT501}
\begin{tabular}{ccc}
\toprule
 & NCT-501 & NCT-501- NC(=S)N \\
\midrule
Structural Formula  & 
  \includegraphics[width=0.15\textwidth]{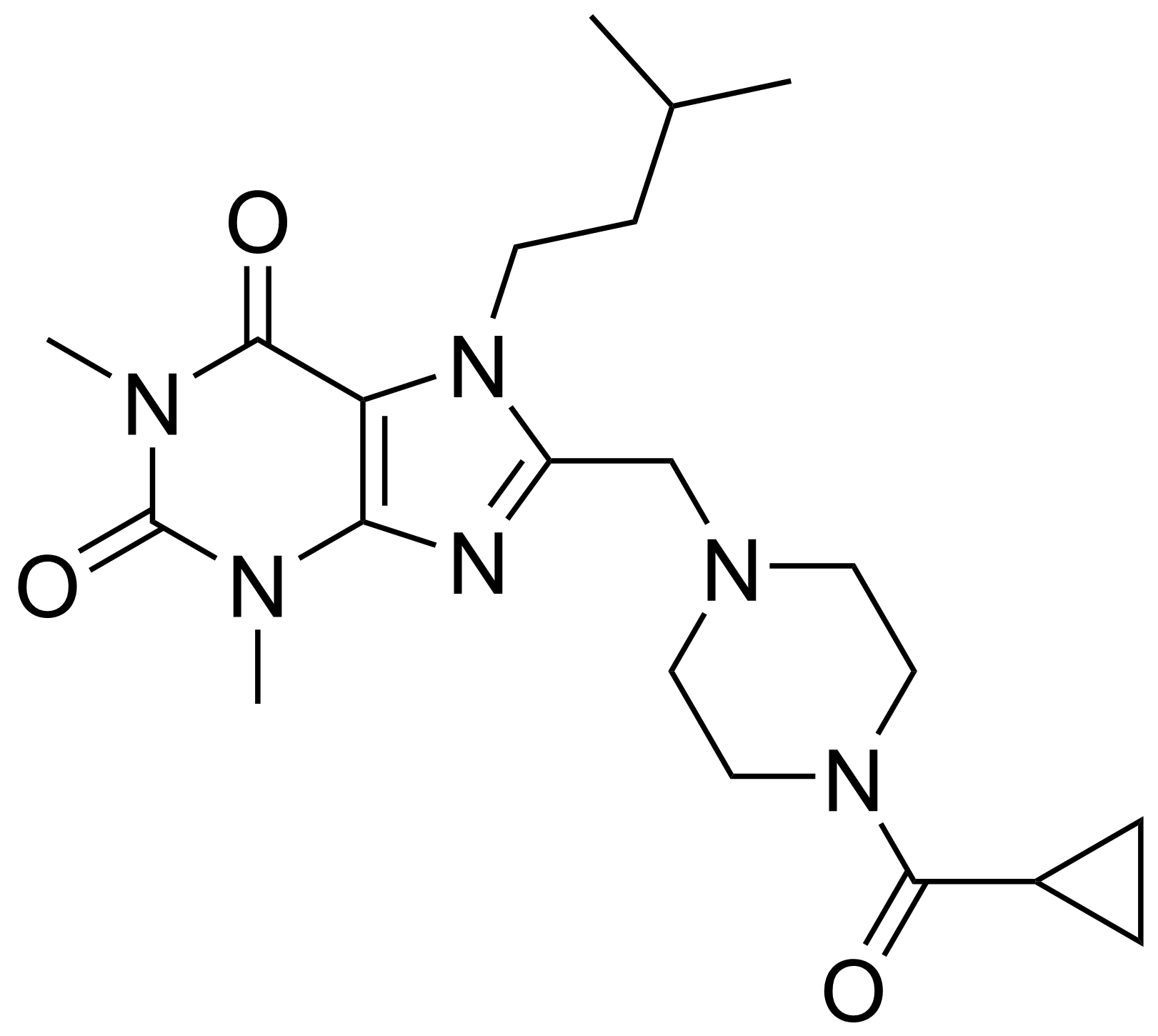} & 
\includegraphics[width=0.15\textwidth]{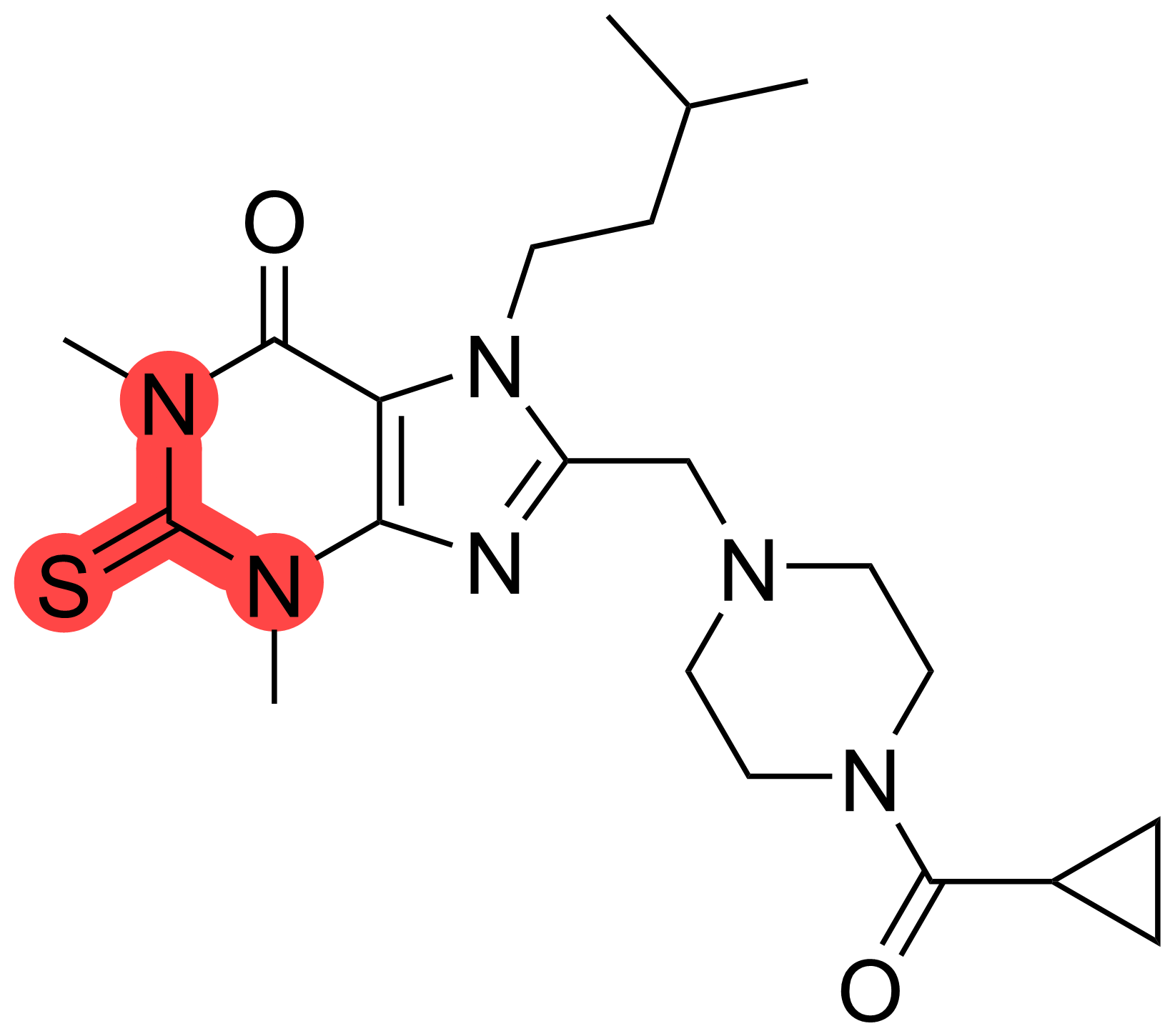} \\ \midrule

Mol. Weight($\leq$500 Da) & 416.522 & 432.583 \\
\midrule
LogP octanol/water($\leq$5) & 1.363  & 2.159 \\
\midrule
HBD($\leq$5) & 0 & 0\\
\midrule
HBA($\leq$10) & 10 & \textbf{9.5} \\ \midrule
Human Oral Absorption \%   & \multirow{2}{*}{72.846} & \multirow{2}{*}{\textbf{82.307}} \\
( >80 high, <25 weak activity) & & \\
\bottomrule
\end{tabular}
\label{tab:nct501}
\end{table}

\begin{table}[H]
\centering
\caption{Physicochemical properties of CM026}
\begin{tabular}{ccc}
\toprule
 & CM026 & CM026-NC(=S)N \\
\midrule
Structural Formula  & 
  \includegraphics[width=0.15\textwidth]{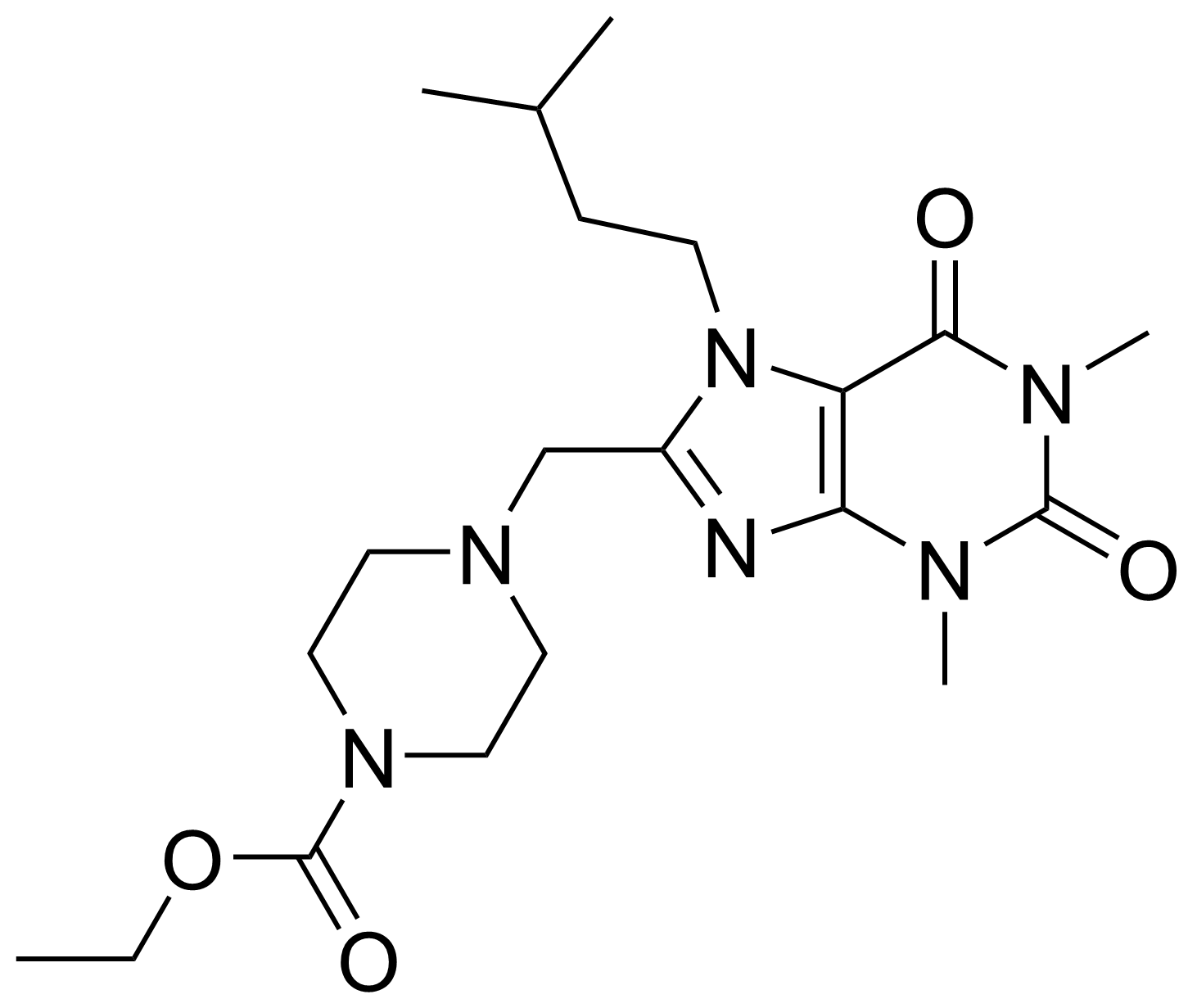} & 
\includegraphics[width=0.15\textwidth]{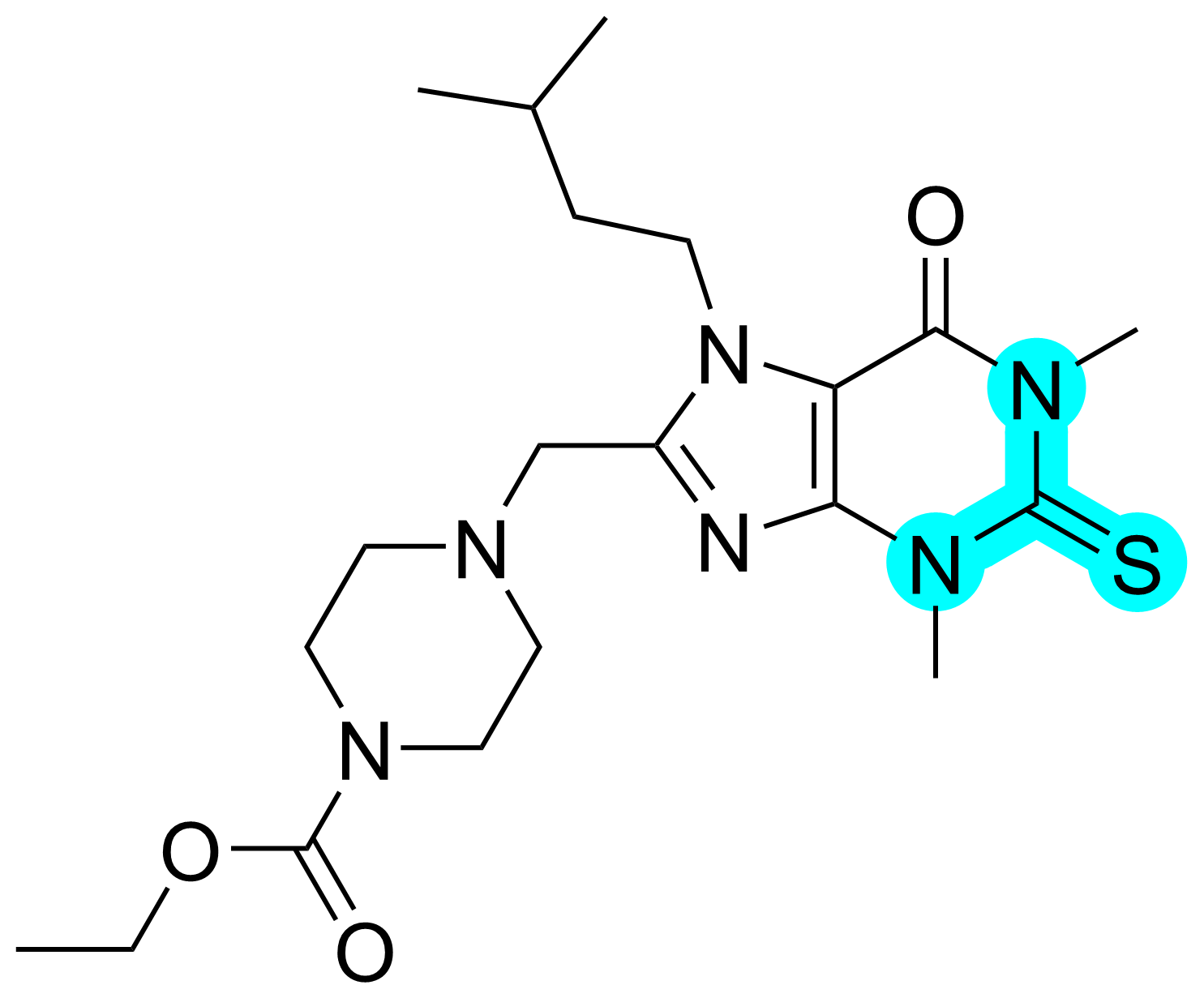}  \\ \midrule

Mol. Weight($\leq$500 Da) & 420.511  & 436.571 \\
\midrule
LogP octanol/water($\leq$5) &2.134  & \textbf{2.954} \\
\midrule
HBD($\leq$5) & 0 & 0\\
\midrule
HBA($\leq$10) & 9.5 & 9 \\ \midrule
Human Oral Absorption \%  &  \multirow{2}{*}{79.551} & \multirow{2}{*}{\textbf{90.170}} \\
\\
( >80 high, <25 weak activity) & & \\

\bottomrule
\end{tabular}
\label{tab:cm026}
\end{table}

\section{Discussion}

The proposed method for identifying key chemical words has several limitations.  
Firstly, the method relies on the use of SMILES notation as the primary representation system for chemicals. While SMILES is widely used and offers readable character strings to represent chemical structures, it has inherent limitations. One significant drawback is its inability to capture and process extensive information regarding the spatial arrangement, bond orientation, and overall 3D structure of molecules. 
Furthermore, during the extraction of language units or tokens, the chemical word identification algorithm struggles to render complete SMILES strings for ring systems, resulting in incomplete words like \texttt{c2ccc(} and \texttt{c1cn}. 
However, these incomplete words still contained valuable chemical information. For example, the word \texttt{c1ccc2[nH]cnc2c1} represents the 1\textit{H}-benzimidazole ring structure. If we were given the SMILES string \texttt{c1ccc2[nH]cnc2c}, we could infer that it contains a 1\textit{H}- imidazole ring based on the presence of \texttt{c2}. However, we would not be able to infer that it contains a 1\textit{H}-benzimidazole ring structure because the ring starting with \texttt{c1} is not completely closed.
To interpret such words on a fragment-based level, we need to have some knowledge of chemistry, such as how rings can be formed and how many bonds an atom can make.
Additionally, there are some words for which no interpretation can be made. For instance, the word \texttt{CC[C@]34C)} seems to indicate isomeric structure, ring, and aliphatic structure, but it is not possible to draw any meaningful chemical inference based on only three carbons.

\section{Conclusion}

In this work, we investigated the  chemical vocabularies generated by three subword tokenization algorithms (BPE, WordPiece, and Unigram) and proposed a novel language-inspired pipeline to select a subset of chemical key words that are strongly associated with protein-ligand binding. The chemical key words are found to be similar across different tokenization algorithms and are specific to each protein or family, often associated with only one protein/family and significantly different from the key words for weak binders.
A detailed study of the literature shows that the chemical key words can designate pharmacophores and/or functional groups for all studied protein families and we also obtained important results that can improve the ADME profiles of the compounds. We view our study as a step towards interpreting the chemistry awareness of chemical words from the perspective of pharmaceutical chemistry and elucidating machine learning models for computational drug discovery, at large.

The success of the proposed pipeline in identifying key chemical substructures encourages prospective studies to discover new pharmacophore groups with machine learning. The proposed algorithm can leverage already available large protein-ligand interaction datasets to annotate more chemicals and propose novel pharmacophore and/or functional groups for protein targets. Such novel groups can guide the discovery of novel drugs and drug-like molecules.

\section{Data and Software Availability}
The LIT-PCBA, BDB, and ProtBENCH datasets are publicly available at the following URLs:
\begin{itemize}
\item LIT-PCBA dataset: \url{https://drugdesign.unistra.fr/LIT-PCBA/}
\item BDB dataset: \url{https://github.com/boun-tabi/chemboost}
\item ProtBENCH dataset: \url{https://github.com/HUBioDataLab/ProtBENCH}
\end{itemize}


\bibliographystyle{unsrtnat}
\bibliography{main}  
\section{Appendix}

\end{document}